\documentclass[journal]{IEEEtran}

\usepackage{amssymb}
\usepackage{multirow}
\usepackage{cite}
\usepackage{float}
\usepackage[cmex10]{amsmath}
\usepackage{ifthen}
\usepackage{etoolbox}
\usepackage{soul}
\usepackage[binary-units=true]{siunitx}
\usepackage{color}
\usepackage[off]{auto-pst-pdf}
\usepackage{psfrag}
\usepackage{epstopdf}
\usepackage{epsfig}
\usepackage[prefix]{nomencl}
\usepackage{algpseudocode}
\usepackage{algorithm}

\usepackage{array} 

\usepackage{lipsum}
\usepackage{etoolbox}
\usepackage{booktabs,colortbl}

\makeatletter
\newcommand{\thickhline}{%
	\noalign {\ifnum 0=`}\fi \hrule height 1pt
	\futurelet \reserved@a \@xhline
}
\newcolumntype{"}{@{\hskip\tabcolsep\vrule width 1pt\hskip\tabcolsep}}
\makeatother

\renewcommand\nomgroup[1]{%
	\item[\bfseries
	\ifstrequal{#1}{P}{ and Constraints}{%
	\\ \ifstrequal{#1}{V}{}{%
	\ifstrequal{#1}{I}{}{}}}]%
}

\makenomenclature

\ifCLASSINFOpdf
   \usepackage{graphicx}
  \graphicspath{{../pdf/}{../eps/}}
  \DeclareGraphicsExtensions{.pdf,.eps}
\else
\fi
\begin{document}
%

\title{A Framework for Frequency Stability Assessment of Future Power Systems: An Australian Case Study}
%

\author{Ahmad Shabir~Ahmadyar~\IEEEmembership{Graduate~Student~Member,~IEEE}, Shariq~Riaz~\IEEEmembership{Graduate~Student~Member,~IEEE},
        Gregor~Verbi\v{c},~\IEEEmembership{Senior~Member,~IEEE},~Archie~Chapman~\IEEEmembership{Member,~IEEE},~and~ 
        David~J.~Hill~\IEEEmembership{Life~Fellow,~IEEE}
\thanks{The authors are with the University of Sydney, Faculty of Electrical and Information Technology, School of Electrical and Information Engineering, Sydney, New South Wales, Australia. E-mail: (ahmad.ahmadyar,~shariq.riaz,~gregor.verbic,~archie.chapman,~david.hill@sydney.edu.au).}
\thanks{Shariq Riaz is also with the Department of Electrical Engineering, University of Engineering and Technology Lahore, Lahore, Pakistan.}

\thanks{David J. Hill is also with the Department of Electrical and Electronic Engineering, The University of Hong Kong, Hong Kong.}}

%
%

\markboth{} 
{Shell \MakeLowercase{\textit{et al.}}: Bare Demo of IEEEtran.cls for Journals}
%



\maketitle

\begin{abstract}
The increasing penetration of non-synchronous renewable energy sources~(NS-RES) alters the dynamic characteristic, and consequently, the frequency behaviour of a power system. To accurately identify these changing trends and address them in a systematic way, it is necessary to assess a large number of scenarios. Given this, we propose a frequency stability assessment framework based on a time-series approach that facilitates the analysis of a large number of future power system scenarios.  We use this framework to assess the frequency stability of the Australian future power system by considering a large number of future scenarios and sensitivity of different parameters. By doing this, we identify a maximum non-synchronous instantaneous penetration range from the frequency stability point of view. Further, to reduce the detrimental impacts of high NS-RES penetration on system frequency stability, a dynamic inertia constraint is derived and incorporated in the market dispatch model. The results show that such a constraint guarantees frequency stability of the system for all credible contingencies. Also, we assess and quantify the contribution of synchronous condensers, synthetic inertia of wind farms and a governor-like response from de-loaded wind farms on system frequency stability. The results show that the last option is the most effective one.         
\end{abstract}
\begin{IEEEkeywords}
Frequency control, future power system security, minimum system inertia, rotational kinetic energy, synchronous condensers, synthetic inertia.
\end{IEEEkeywords}


%
\IEEEpeerreviewmaketitle

\section{Introduction} \label{intro}
\IEEEPARstart{F}{uture} power system~(FPS) security will be challenged by the increasing penetration of non-synchronous renewable energy sources~(NS-RES). With high penetration levels of NS-RES, power system's dynamic complexity increases, which makes it very difficult to define the worst case scenarios used for system stability assessment. Furthermore, with a high level of uncertainty related to NS-RES, even the process of defining the worst case scenarios becomes a challenging task. Therefore, the traditional way of assessing system stability, where the worst case scenarios are usually defined, may no longer be valid. As a result, using the current widely accepted criteria for defining critical operation points and contingencies may fail to identify all critical conditions that might threaten power system stability. This, particularly, has a profound impact on system frequency stability where the increased penetration of NS-RES reduces the total inertia, as well as the governor response of the system. This, in turn, has adverse effects on the system frequency control, and can significantly change the system frequency behaviour, as characterised by the rate of change of frequency~(RoCoF) and the frequency nadir~\cite{EirGridandSONI2010, Miller2011a, Miller2014, Doherty2010}. 
\par Therefore, with a high penetration of NS-RES, system frequency control becomes a challenging task, and needs to be dealt with in a systematic way. Thus, in this paper, we propose a frequency stability assessment framework to evaluate the frequency stability of a system with high penetration of NS-RES for a large number of scenarios. We utilise the proposed methodology to assess the frequency stability of the Australian FPS for a large number of scenarios by considering different penetration of NS-RES and prosumers\footnote{Consumers equipped with distributed generation (e.g. rooftop-PV) and battery storage~(BS) are referred to \emph{prosumers} (\textbf{pro}ducer - con\textbf{sumer}).}. Further, we perform sensitivity analysis to study the impact of different load model and inertia location on system frequency stability. Finally, we use some of conventional and non-conventional resources to improve system frequency response, and quantify their contribution on system frequency control.    
\par Studies have shown that with high penetration of NS-RES, it might be very difficult to maintain system frequency within the stable limits~\cite{Miller2014, Miller2011a, ERCOT2013, Doherty2010, EirGridandSONI2010, AEMO2016a, TheEirGridGroup2015}. In~\cite{Miller2014}, the frequency stability of the US Western Interconnection was assessed. The results show that a \SI{50}{\percent} non-synchronous instantaneous penetration~(NSIP) deteriorates the frequency response of the system compared to the current situation; however, it is not considered as an immediate issue provided that adequate fast frequency responsive resources are available. In contrast, EirGrid in Ireland has introduced a maximum of \SI{50}{\percent} NSIP to maintain the system frequency response within the permissible bounds~\cite{TheEirGridGroup2015}. Although the above studies give some insight into the effect of high penetration of NS-RES on system frequency stability, it is difficult to make a general conclusion. Nonetheless, studies have shown that system frequency stability is an especially serious issue for small and weakly interconnected systems, such as Ireland grid~\cite{EirGridandSONI2010} and the Australian National Electricity Market~(NEM). Hence, the Australian Energy Market Operator~(AEMO) has identified system frequency control as an immediate challenge for FPS security~\cite{AEMO2016a}, mainly because the transmission network is very long, and areas can be disconnected. 
\par Currently, in the NEM, eight frequency control ancillary service~(FCAS) markets conjointly operate with other spot markets in real time~\cite{AEMO2015}. These FCAS markets provide primary frequency support with three different speeds of response - \SI{6}{\second}, \SI{1}{\minute}, and \SI{5}{\minute}. Hence, the primary frequency control might not be considered as an immediate issue for the FPSs provided that a similar model is implemented. On the other hand, a higher RoCoF, which is a consequence of low system inertia, is considered as an immediate challenge~\cite{AEMO2016a,ERCOT2013}. Indeed, low system inertia was identified as one of the main reasons for the South Australia's~(SA) blackout on 28 September 2016~\cite{AEMO2016d}. After separation of the SA system from rest of the NEM, inertia was very low in the SA system, which resulted in a considerably fast RoCoF; consequently, the system frequency dropped below \SI{47}{\hertz} before the activation of the under frequency load shedding~(UFLS) scheme. 
\par Note that a higher RoCoF leaves insufficient time for activation and deployment of primary frequency response, and it might trigger generators RoCoF protection, which would result in tripping of additional synchronous generators~(SGs). Further, with the increased RoCoF, the maximum torque and mechanical stress increases in the machines. In some situations, with a RoCoF of $\SI[]{-1}{\hertz\per\second}$, an under-excited machine's rotor cannot follow the fast speed reduction of the grid, so it loses its opposed force and speeds up, which would eventually result in pole slip~\cite{DNVKEMAEnergy2013}. It has also been reported that fast frequency events can have detrimental effects on combustion turbines because of potential turbine combustor lean-blowout~\cite{NERC2008}.  
\par Therefore, with a high penetration of NS-RES, it is crucial to maintain system frequency within its stable limits. To deal with this issue, some studies have proposed a maximum NSIP limit, which requires a certain portion of power to be produced by the SGs at all times~\cite{EirGridandSONI2010, OSullivan2014}. Similar approaches have been proposed in~\cite{Doherty2010} and \cite{Ahmadi2012}, where a frequency control security constraint was integrated in the market dispatch model. However, before integrating any frequency control security constraint in the market dispatch model, it has to be defined precisely, and the limits for this constraint have to be identified accurately. To do this in a systematic way, it is vital to have a suitable framework to precisely assess the frequency stability of FPSs, and determine the maximum NSIP that a system can operate without violating its frequency stability limits.

\par With such an objective, this paper proposes a frequency stability assessment framework based on time-series analysis in Section~\ref{Section_FSAF}. Then, in Section~\ref{FS-A-NEM}, we use the proposed framework to perform scenario based sensitivity analysis, which allows us to accurately assess the frequency stability of the Australian FPS for all credible contingencies~(CCs; e.g. loss of largest in-feed generator). By doing this, we can identify the highest NSIP ranges that the system can withstand without violating its frequency stability limits. In Section~\ref{Slution}, we employ the proposed framework to incorporate a RoCoF based dynamic inertia constraint into the market dispatch model to improve frequency stability of the system with high penetration levels of NS-RES. Further, we use some of the conventional, such as synchronous condensers~(SC)~\cite{Kundur1994}, and non-conventional resources, such as wind farm's (WF's) synthetic inertia~\cite{Ekanayake2004}, and a governor-like response from WFs to improve frequency response of the system. By doing this, we can evaluate and quantify the contribution of these resources on system frequency control, as well as identify the most effective option for system frequency control. Note that the WFs' governor-like response can be achieved by initially curtailing wind energy and operating the WFs in a de-loaded~(suboptimal) mode, and changing their operation mode from sub-optimal to optimal in the inception of an event~\cite{Ahmadyar2017}. Finally, conclusions are presented in Section~\ref{concl}. 
\begin{algorithm}[t]
	\caption{FPS Frequency Stability Assessment Framework} \label{FPS-FSAF}
	\textbf{Inputs}: Network data, generation data, ancillary service requirements~(e.g. spinning reserve), wind, solar and demand traces for each scenario $s \in \mathcal{S}$ in the studied year.
	\begin{algorithmic}[1] \small
		\For{$s \gets 1, \left| \mathcal{S} \right|$}
		\For{$t \gets 1, \left| \mathcal{T} \right|$}
		\State{Market simulation (generation dispatch);}
		\State{Identify credible contingencies;}
		\State{Load-flow analysis;}
		\EndFor
		\For{$c \gets 1, \left| \mathcal{C} \right|$}
		\For{$t \gets 1, \left| \mathcal{T} \right|$}
		\State{Frequency stability assessment by;}
		\State{Considering all the credible contingencies;}
		\EndFor
		\EndFor
		\EndFor
	\end{algorithmic}
	\textbf{Outputs}: Frequency stability indices (i.e. minimum RoCoF and frequency nadir) for each time slot $t \in \mathcal{T}$, for each sensitivity case $c \in \mathcal{C}$, and for each scenario $s \in \mathcal{S}$.
\end{algorithm}
\vspace{-1em}
\section{Frequency Stability Assessment Framework and Case Studies} \label{Section_FSAF}
\par Our framework is based on the stability scanning framework proposed in~\cite{Marzooghi2014,Liu2017}. Algorithm~\ref{FPS-FSAF} summarises the proposed framework. Notice that it not only captures the most relevant aspects of power system operation that alter frequency response of the system following a CC, but also defines and identifies the CCs for large number of scenarios using time-series approaches.
\vspace{-1.3em} 
\subsection{Inputs and Scenarios} \label{SandI} 
\par
\begin{figure}
	\begin{center}
		\includegraphics[width=4.5cm, keepaspectratio]{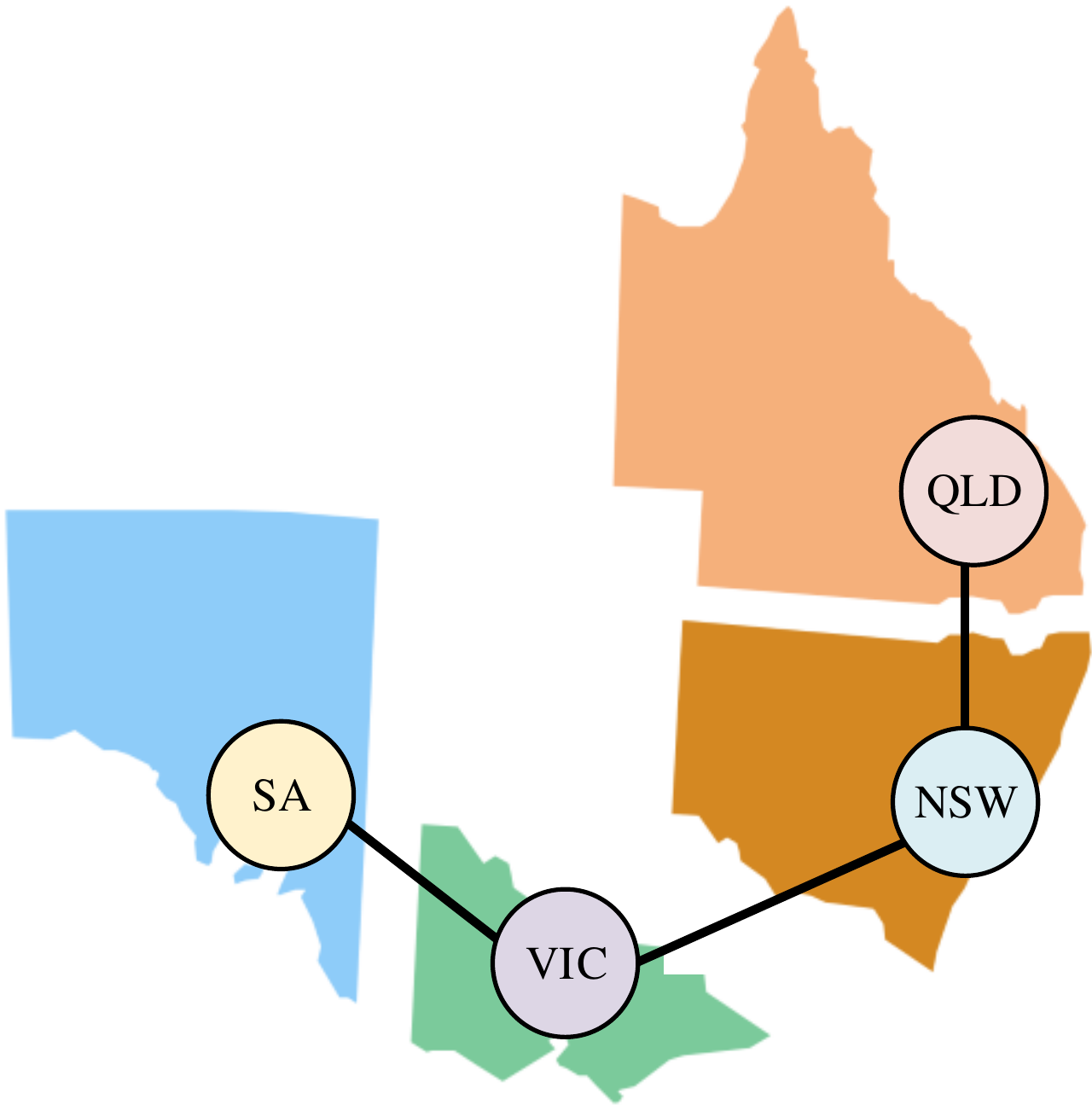}
\vspace{-1em}
	\end{center}
	\caption{Simplified network diagram of the NEM.}
\vspace{-0.5em}	
	\label{SLD}
\end{figure}

 Unlike previous deterministic studies where a few specific operation points were used to assess the frequency stability of a system~\cite{Miller2014}, our framework utilises a large number of scenarios, which is crucial for identifying principal issues related to system frequency stability. For extensive time-series studies proposed in this paper, use of a simplified network model would be sufficiently accurate and computationally  efficient~\cite{Doherty2010}. Although the simplified model provides an approximate result, it would be a good representation of qualitative trend in the system dynamic behaviours. Furthermore, the Australian NEM has one of the longest transmission networks~(i.e. more than \SI{5000}{\kilo\meter} long), in which four states are weakly connected~\cite{Gibbard2010}. Considering these, we use a simplified four-area network model of the NEM as shown in Fig.~\ref{SLD}, where each state is represented by one area~(i.e. Queensland~(QLD), New South Wales~(NSW), Victoria~(VIC) and SA). The inter-state transmission lines are augmented accordingly to accommodate the new generation and demand for the year 2040~\cite{AEMO2013}. As the base scenario, we use the current NEM generation portfolio, summarised in~Table~\ref{GenPort}\footnote{http://www.nem-watch.info/}, with approximately~\SI{10}{\percent} non-synchronous annual penetration~(NSAP)~(i.e. \SI{10}{\percent} of annual energy is supplied by NS-RES predominantly WFs). Then, we incrementally retire the coal fired generators and increase the NSAP by~\SI{10}{\percent} up to~\SI{90}{\percent}. To do this, we use the Australian~\SI{100}{\percent} renewable studies as guidelines~\cite{Riesz2016, AEMO2013, Institute2010}; this will result in a total of nine different scenarios; $\left| \mathcal{S} \right| = 9$. In order to perform a comprehensive study, each scenario is analysed for one year at hourly resolution, $\left| \mathcal{T} \right| = 8760$, to capture a wide range of intra day and inter seasonal demand and generation variations. The hourly demand, wind and solar traces for each scenario are extracted from the Australian National Transmission Network Development Plan (NTNDP)~\cite{AEMO2015a}. 
\vspace{-0.5em}
\begin{table}[]
	\centering
	\caption{The assumed large-scale generation portfolio of NEM; base scenario with~\SI{10}{\percent} annual energy from NS-RES.} 
\vspace{-0.8em}
	\label{GenPort}
	\begin{tabular}{c||c|c|c|c}
		\toprule \hline
		Plants & \begin{tabular}[c]{@{}c@{}}No.of \\ Unit\end{tabular} & \begin{tabular}[c]{@{}c@{}}Annual Energy\\ (TWh)\end{tabular} & \begin{tabular}[c]{@{}c@{}}Capacity\\ (GW)\end{tabular} & \begin{tabular}[c]{@{}c@{}}Inertia Constant\\ $H$ (s)\end{tabular} \\ \hline \hline
		Hydro     & 7  & 20.2  & 2.3  & 4 \\ \hline
		Coal      & 77 & 157.6 & 39.4 & 6 \\ \hline
		CCGT      & 5  & 0.9   & 1.7  & 6 \\ \hline
		OCGT      & 12 & 1.4   & 3.6  & 6 \\  \hline
		WF        & -  & 17.8  & 5.9  & 0 \\\hline \bottomrule
	\end{tabular}
\vspace{-1.5em}
\end{table}
\vspace{-0.8em}
\subsection{Market Simulation (Line 1-6,~Algorithm~\ref{FPS-FSAF})}
\par In restructured power systems, market dynamics can affect system operation, and consequently, system frequency stability~\cite{Alvarado2001}. Furthermore, system frequency dynamics can be influenced by several factors as follows~\cite{Kundur1994}:
\begin{equation} \label{rocof_formula}
\frac{df}{dt}=\frac{f_\text{0}}{2I_\text{s}}p^\text{cc}-\frac{f_0}{2I_\text{s}D_\text{load}} f,
\end{equation}
where $f_\text{0}$ is the reference frequency of the system prior an incident, $p^\text{cc}$ is size of the CC, $D_\text{load}$ denotes the frequency damping of the system load, $f$ is the system frequency, and $I_\text{s}$ represents the level of system synchronous inertia, given by:
\begin{equation} \label{Equation_Inetia}
I_\text{s} = \sum\limits_{i=1}^{N} H_{\text{g},i} S_{\text{B},i}, 
\end{equation}
where $H_{\text{g},i}$ and $S_{\text{B},i}$ are inertia constant, and rated MVA of generator $i$, respectively, and $N$ is the number of online SGs in the system. Note that the above parameters as well as system governor response, and system primary reserve would change from one period to the following; therefore, to capture the effect of market dynamics and other changes on system frequency stability, we run the market dispatch model. The market dispatch model is based on a modified unit commitment (UC) problem that aims to fulfil the demand requirement while minimising generation cost by considering fixed, start-up, shut-down and fuel cost of all generators. In addition, the dispatch decisions are constrained by the minimum thermal stable limits, ramp rates, minimum up/down time of SGs, renewable resource availability and thermal limits of transmission lines~\cite{Riaz17}. In the market dispatch model, it is assumed that the conventional SGs, such as coal fired power plants, open cycle gas turbines (OCGT) and combined cycle gas turbines (CCGT) bid at their respective short-run marginal cost (SRMC), estimated using the predicted fuel price, thermal efficiency, and variable operation and maintenance~(O\&M) cost in 2040~\cite{Tasman09}; while, the NS-RES generators~(e.g. WFs and PV plants) bid with zero SRMC. Notice that the NS-RES would push the conventional SGs out of the merit order. This, in turn, can significantly reduce the size and the operation point of CCs~(e.g. largest in-feed units). Therefore, during the market simulation, we identify the size and operating point of the CCs using equation\footnote{Due to high simulation burden, we consider only the largest in-feed unit as a CC and find its operation point. For more detail studies, we can monitor the in-feed power at every node to find the node with the highest in-feed power, and consider it as a CC.}~(\ref{eq:cntgcy}).
\begin{equation}
p_t^\text{cc}=\mathop{\operatorname{max}}\,(p_{g,t}) \quad g\in \mathcal{G^\text{SG}} \text{,}	\label{eq:cntgcy}
\end{equation} 
where, $p_{g,t}$ represents the dispatch level of generator $g$ during time period $t$, and $\mathcal{G^\text{Syn}}$ is the set of all SGs. Once the dispatch results for a specific hourly scenario are generated and the CCs are identified, we perform load flow analysis, which provides the initial condition for dynamic simulations. 
\begin{figure}
	\begin{center}
		\includegraphics[width=7.8cm, keepaspectratio]{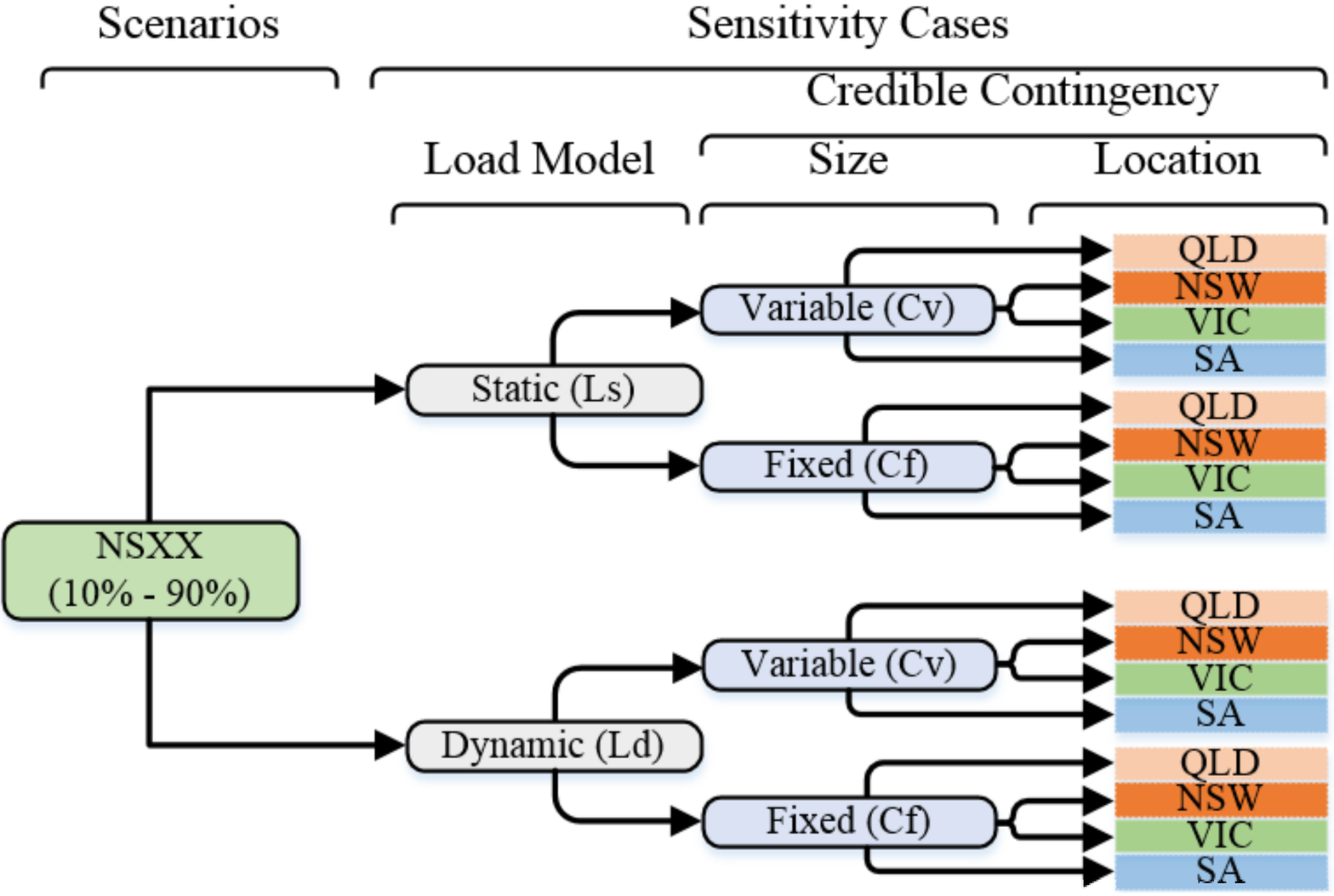}
	\end{center}
	\caption{A summary of scenarios and sensitivity cases, NSXX is the NSAP and varies from \SI{10}{\percent} to \SI{90}{\percent}.}
\vspace{-1.5em}
	\label{SandS}
\end{figure}
\vspace{-1.2em}
\subsection{Sensitivity Cases and Frequency Stability Assessment \\
	(Line 7-13,~Algorithm~\ref{FPS-FSAF})}  \label{SAandSC}
\par For a comprehensive frequency stability assessment, we consider the sensitivity of some of the most relevant parameters given in~(\ref{rocof_formula}) including the load model, as well as the size and the location of contingencies as shown in Fig.~\ref{SandS}. The static load model~(\emph{Ls}) is a constant impedance, current and power~(ZIP) model~\cite{IEEETaskF1993}; whereas, the dynamic load model~(\emph{Ld}) consists of~\SI{40}{\percent} induction machines~\cite{Hiskens1995}, and~\SI{60}{\percent} ZIP. For the size of the variable contingency~(\emph{Cv}), we use the values obtained from (\ref{eq:cntgcy}); whereas, for the size of the fixed contingency~(\emph{Cf}), we use the size of the largest in-feed generator that currently operates in the NEM, i.e.~\SI{666}{\mega\watt}~\cite{Gibbard2010}. We consider the above cases in all areas; namely, QLD, NSW, VIC and SA. So, considering the number of scenarios~$\left| \mathcal{S} \right| = 9$, and sensitivity cases (2x2x4)=16, we will assess a total of 144 cases in a year;\textbf{ $\left| \mathcal{C} \right| = 144$}, with \textbf{$\left| \mathcal{T} \right| = 8760$} in a year; resulting in \textbf{\num{1261440}} simulation runs. Based on Fig.~\ref{SandS}, we use the following convention to present the results; for the form \emph{NSxx-Ls/d,Cv/f,Location-Metre}: \textbf{NSxx} is the NSAP, i.e.~\SI{10}{\percent} to \SI{90}{\percent}; \textbf{Ls/d} represents the load model, i.e. \textbf{Ls} and \textbf{Ld}; \textbf{Cv/f} shows the contingency size, i.e. \textbf{Cv} and \textbf{Cf}; \textbf{Location} provides information regarding the location of the CC; and \textbf{Metre} is the bus where the measurements were taken. An example is give as follows:
\begin{figure}
	\begin{center}
		\includegraphics[width=8.5cm, keepaspectratio]{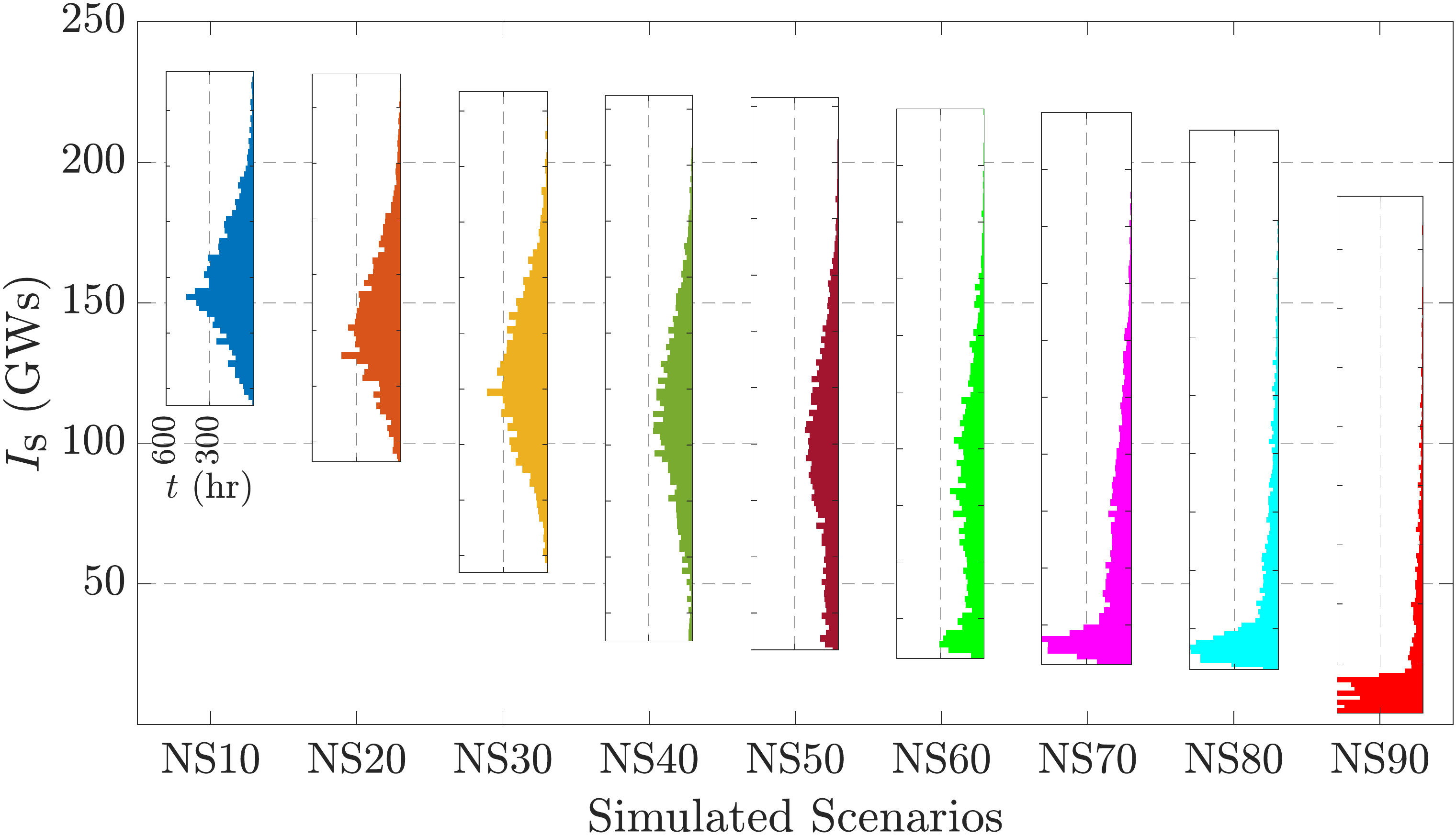}
\vspace{-1em}
	\end{center}
	\caption{Histograms of system synchronous inertia for all scenarios; for consistency all histograms are plotted on the same scale.}
\vspace{-1.2em}
	\label{Inertia_NS}
\end{figure}
\vspace{-0.4em}
\begin{itemize}
	\item \emph{NS40-LdCvNSW-SA}; the NSAP is \SI{40}{\percent}; dynamic load model is used; contingency size is obtained from equation~(\ref{eq:cntgcy}); contingency location is NSW; and the measurement is taken from the SA's bus. 
\end{itemize}
\vspace{-1.2em}
\subsection{Outputs} \label{Ouput}
\par The proposed framework provides a set of outputs for every simulated hour; namely, the minimum RoCoF and frequency nadir. We use these output indicators to estimate the critical range of NSIP that the Australian FPS can accommodate from a system frequency stability point of view. The NSIP for every period $t$ is defined as follows:
\begin{equation} \label{Equation_NSIP}
\text{NSIP} = \frac{p_t^\text{NS-RES}}{p_t^\text{NS-RES}+p_t^\text{SGs}} \text{,}	
\end{equation}     
where $p_t^\text{NS-RES}$ and $p_t^\text{SGs}$ are the power contribution of NS-RES and SGs, respectively.
\vspace{-0.7em}
\section{Frequency Stability Assessment Results}\label{FS-A-NEM}
\par The framework proposed in Section~\ref{Section_FSAF} is used for the frequency stability assessment of the Australian FPS. All SGs are modelled with their governor, excitation system and power system stabilisers~\cite{Gibbard2010, Ahmadyar2016}. Also, all WFs are modelled as Type~IV wind turbine generators, and all utility PV plants are modelled as full converter interface generators.
\begin{figure}
	\begin{center}
		\includegraphics[width=8.5cm, keepaspectratio]{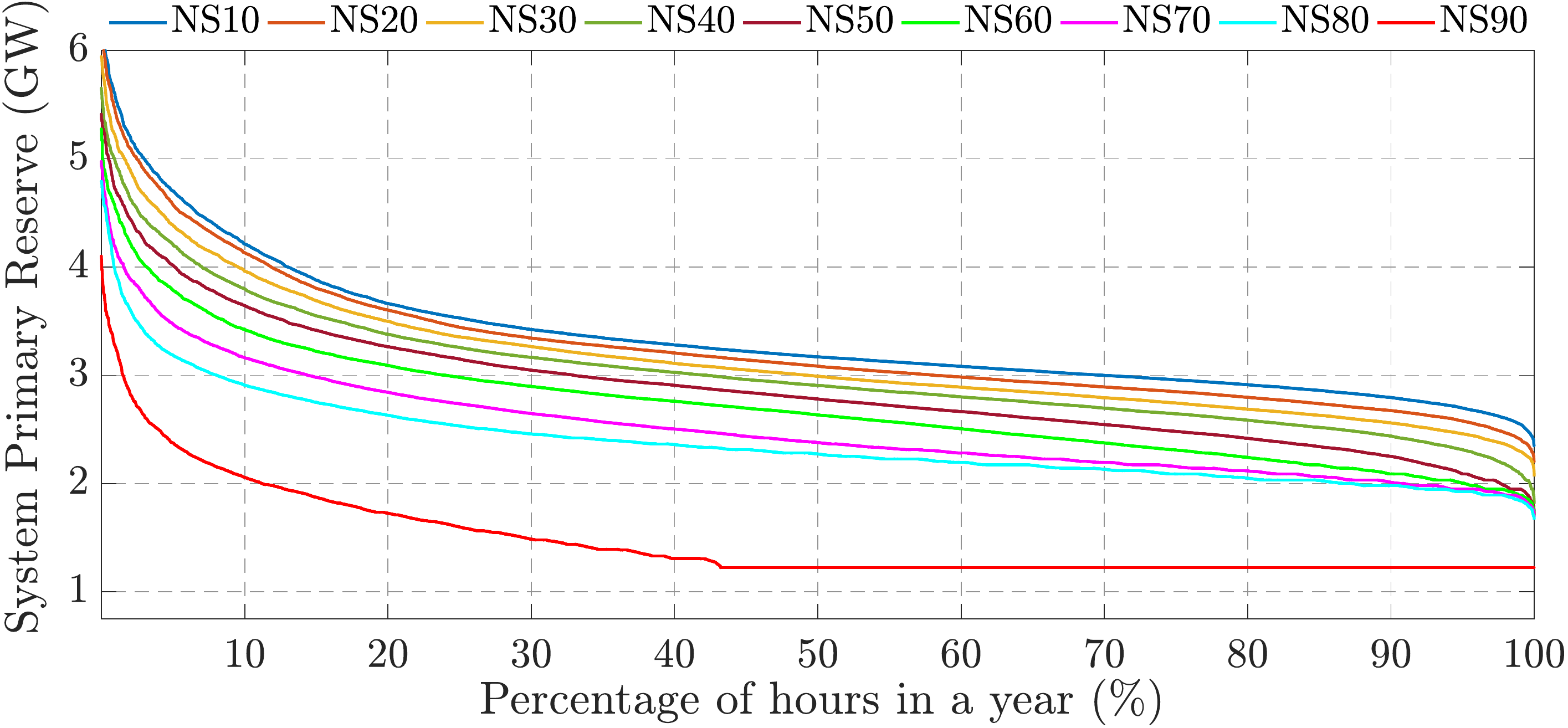}
\vspace{-1.5em}
	\end{center}
	\caption{System primary reserve for all scenarios; $P^{\text{max}}_{g}(t) - P_{g} (t), \; g\in \mathcal{G^\text{Syn}}$.}
\vspace{-1.5em}
	\label{Headroom}
\end{figure}
\vspace{-0.8em}
\begin{figure}
	\begin{center}
		\includegraphics[width=8.5cm, keepaspectratio]{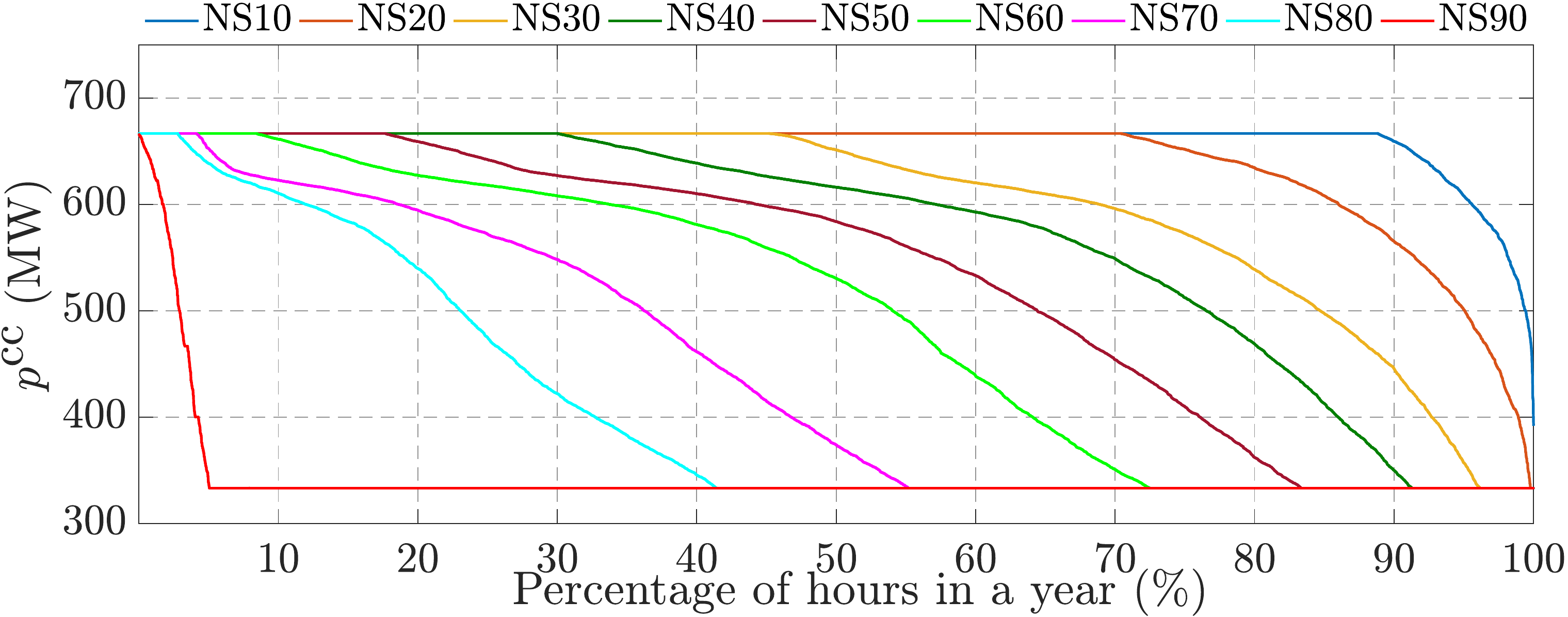}
\vspace{-1em}
	\end{center}
	\caption{Size of credible contingency~(\emph{Cv}) based on Algorithm~\ref{FPS-FSAF}.}
\vspace{-1.5em}
	\label{Contingency}
\end{figure}
\vspace{-0.5em}
\subsection{Market Dispatch Results} \label{Marekt_Dispatch_Results}
\par We perform UC simulation for the nine scenarios introduced in Section~\ref{Section_FSAF}. First, we examine the generation mix generated by solving the nine UC scenarios. To deal with the unpredictability of load and generations, we considered a minimum of \SI{10}{\percent} spinning reserve from SGs for scenarios \emph{NS10} to \emph{NS80} in each region. With such a constraint, it was infeasible to increase the NSAP to above \SI{83}{\percent}; therefore, in scenario~\emph{NS90} we dropped the spinning reserve constraint. Notice that behind scenario~\emph{NS80} the system is predominantly renewable based with some biofuel fired OCGTs in each region that provide operational flexibility for the system~\cite{Riesz2016}; hence, scenario~\emph{NS90} can be considered as a~\SI{100}{\percent} renewable based network with a generation mix of WFs, Hydro, concentrated solar power~(CSP), PV-plants and OCGTs. Second, we evaluate the impact of high penetration of NS-RES on the amount of available synchronous inertia and system primary reserve, which can be directly linked to the RoCoF and frequency nadir~\cite{Kundur1994}. As illustrated in Fig.~\ref{Inertia_NS}, with the increasing penetration of NS-RES, not only the system synchronous inertia reduces, but it also becomes more time variant. For instance, in scenario~\emph{NS10}, the synchronous inertia varies between \num{114}$\,$GWs and \num{234}$\,$GWs with a mean value of \num{159}$\,$GWs; whereas, in scenario~\emph{NS90}, the variation is between \num{4}$\,$GWs and \num{178}$\,$GWs with a mean value of \num{18}$\,$GWs. The system primary reserve follows a similar trend, as shown in Fig.~\ref{Headroom}. The combined reduction of synchronous inertia as well as system primary reserve can have significant detrimental impacts on system frequency stability, which will be analysed in the next subsection. Third, we evaluate the changes in the size of CCs. As shown in Fig.~\ref{Contingency}, there is a significant discrepancy among the sizes of CCs in different scenarios, e.g. in scenario~\emph{NS10}, the size of CC falls below \SI{666}{\mega\watt} approximately \SI{10}{\percent} of the time; whereas, in scenario~\emph{NS80}, this time increases to more than \SI{95}{\percent}. This might have a positive impact on system frequency stability. Therefore, it is very important to consider these changes and assess their implication on system frequency stability. Finally, we use the UC results to perform load flow analysis. The load flow results are used as the initial conditions for the frequency stability assessment of the system in the following subsection.              
\begin{figure}
	\begin{center}
		\includegraphics[width=8.5cm, keepaspectratio]{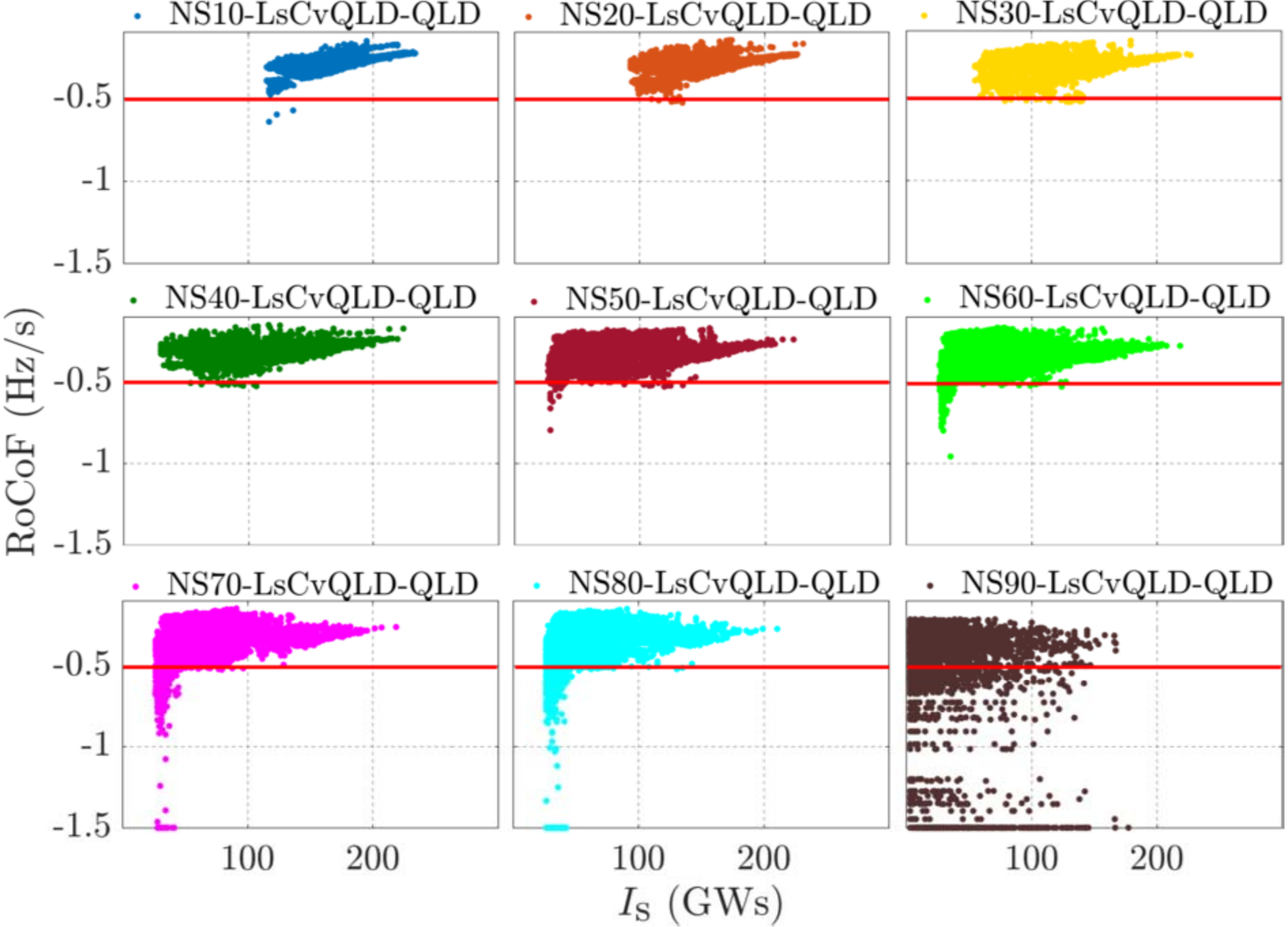}
\vspace{-1.2em}
	\end{center}
	\caption{Minimum RoCoF following the loss of CC for the simulated year; red line shows the critical RoCoF.}
\vspace{-1.5em}
	\label{Fig_RoCoF_all}
\end{figure}
\vspace{-1.2em}
\subsection{Frequency Stability Results} \label{stability_Results}
\par
\begin{figure}
	\begin{center}
		\includegraphics[width=7.5cm, keepaspectratio]{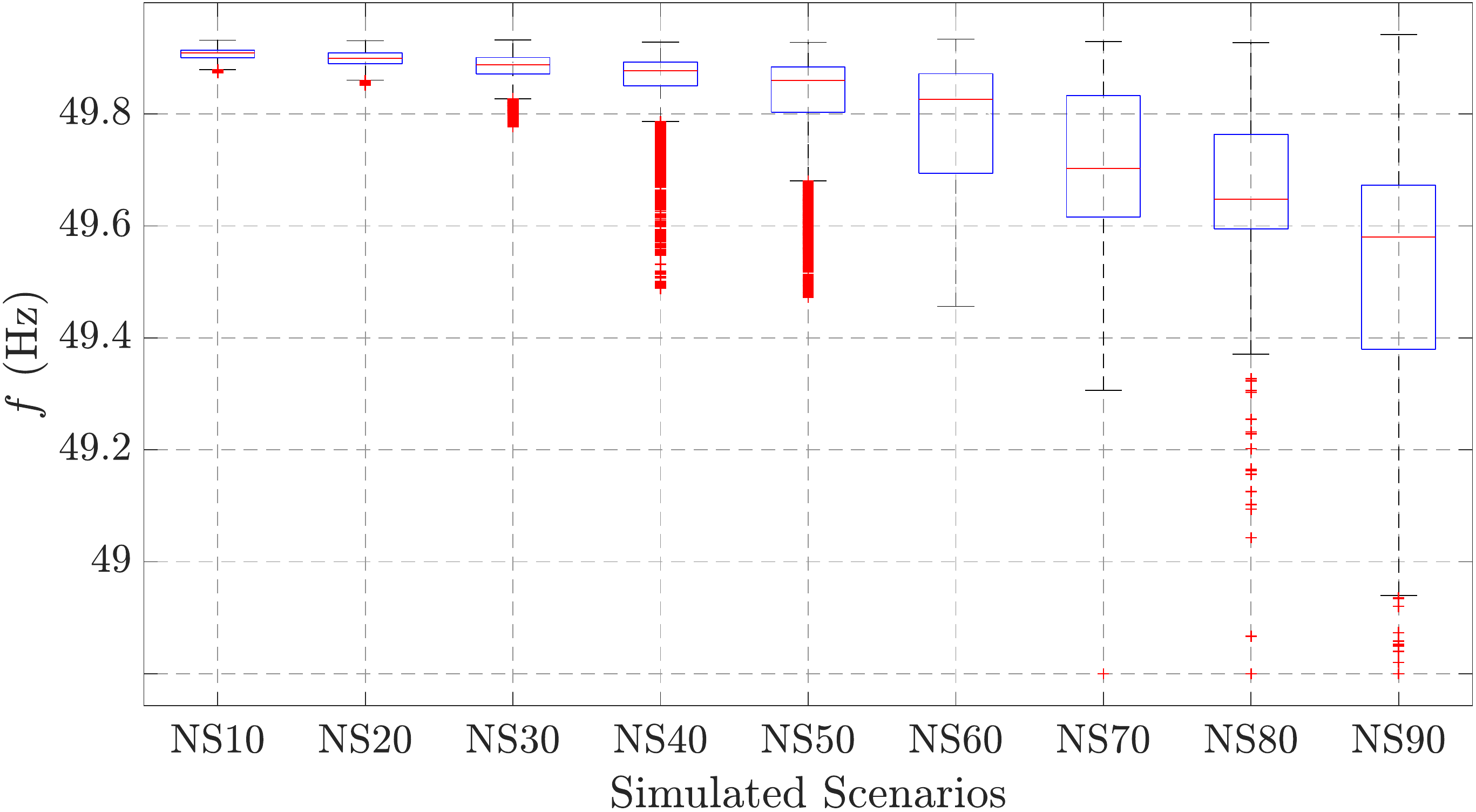}
\vspace{-1.2em}
	\end{center}
	\caption{Frequency nadir following loss of CC for Cases \emph{NSxx-LsCvQLD-QLD}.}
\vspace{-1.5em}
	\label{DF_all}
\end{figure}

 We analyse the frequency behaviour of the system for the first \SI{50}{\second} following a CC for all the cases introduced in Section~\ref{Section_FSAF}. This will allow us to quantify the impacts of reduced inertia, and primary reserves on system frequency stability; namely RoCoF and frequency nadir. Fig.~\ref{Fig_RoCoF_all} summarises the minimum RoCoF for all scenarios of \emph{NSxx-LsCvQLD-QLD}. Observe that how the minimum RoCoF decreases with the reduced amount synchronous inertia, and for the NSAP of \SI{60}{\percent} and above the minimum RoCoF starts violating the critical RoCoF~(i.e. $\SI[]{-0.5}{\hertz\per\second}$~\cite{EirGridandSONI2010}). The frequency nadir follows a similar trend as shown in Fig.~\ref{DF_all}; where above \SI{60}{\percent} NSAP, the number of hours that the frequency nadir drops below \SI{49.5}{\hertz}~(i.e. minimum allowable frequency during a contingency~\cite{AEMO2016}) increases significantly, and for scenario \emph{NS90} even the mean value of frequency nadir is close to \SI{49.5}{\hertz}. Notice that the exclusion of spinning reserve in scenario \emph{NS90} resulted in low system inertia. Also, in some situations, despite a high levels of system inertia, some regions had very low or even zero inertia. As a result, in these situations even with high level of system inertia, we have noticed a high RoCoF for some regions as shown in Fig.~\ref{Fig_RoCoF_all}, Case~\emph{NS90-LsCvQLD-QLD}. This demonstrates the importance of inertia location on system frequency stability. We will return to this point in Section~\ref{Slution} when discussing the technical options for improving system frequency stability.               
\par Notice that how the reduced amount of synchronous inertia and primary reserve in Figs.~\ref{Inertia_NS}~and~\ref{Headroom} affect the RoCoF and the frequency nadir of the system in Figs.~\ref{Fig_RoCoF_all}~and~\ref{DF_all}, respectively. Although inertia level is one of the most relevant indicators of system frequency characteristic, it is not the sole parameter that influences the frequency behaviour of a system. There are other factors that need to be considered while defining the critical level of system inertia as discussed in Section~\ref{Section_FSAF}. For instance, observe in Fig.~\ref{Fig_RoCoF_all}, for scenarios~\emph{NS60},~\emph{NS70} and \emph{NS80}, that for a wide range of inertia level~(i.e. $I_\text{s}=\SI{30}{\giga\watt\second}~\text{to}~\SI{50}{\giga\watt\second}$) the RoCoF might violate its limit in some hours but not in the others. Therefore, based on the above results, it would be impractical to define a single accurate critical inertia level, and consequently, NSIP. Thus, to accurately estimate the critical NSIP of the system from the system frequency stability point of view, we analyse the impact of different parameters that influence the frequency behaviour of the system; namely, load model, CC size, CC location and prosumers in the following subsections. 
\subsubsection{The Impact of the Load Model}
\par The effect of the load model on the system frequency stability performance is significant. As an example, Fig.~\ref{LdLs_CfCv_CLocation}a illustrates the RoCoF considering static load model~(\emph{NS80-LsCvVIC-VIC}) and dynamic load model~(\emph{NS80-LdCvVIC-VIC}). Observe that with the \emph{Ls}, the RoCoF deteriorates compared to \emph{Ld}. Overall, it was observed that the impact of load model is more evident with a high NSIP and load level, e.g. Fig.~\ref{LdLs_CfCv_CLocation}a, $t=\SI{36}{\hour}$. This is because with a high NSIP and load level, the system inertia decreases, while the frequency damping of the system load~($D_\text{load}$) increases. Therefore, the term including $D_\text{load}$ in equation~(\ref{rocof_formula}) becomes more dominant; as a result, the contribution of load dynamic becomes more significant.   
\subsubsection{The Impact of Contingency Size}
\par The impact of contingency size is evidently significant on system frequency stability indices. For instance, observe in Fig.~\ref{LdLs_CfCv_CLocation}b that the RoCoF is notably better in Case~\emph{NS80-LsCvVIC-VIC}, where the CC size was identified during the market simulation, compared to Case~\emph{NS80-LsCfVIC-VIC}, where a fixed CC size was assumed. Observe how the size of CC in Fig.~\ref{Contingency} affects the RoCoF in Fig.~\ref{LdLs_CfCv_CLocation}b. As shown in Fig.~\ref{Contingency} for scenario~\emph{NS80} only less than \SI{5}{\percent} of the time the CC size obtained from Algorithm~\ref{FPS-FSAF}~(\emph{Cv}) is the same as the assumed CC size~(\emph{Cf}); while, in about \SI{60}{\percent} of the time the \emph{Cv} size can be as small as half size of the \emph{Cf}. As shown in equation~(\ref{rocof_formula}) the size of CC is directly proportional to RoCoF. As a result, the RoCoF is considerably better in Case~\emph{NS80-LsCvVIC-VIC}, with variable contingency size compared to Case~\emph{NS80-LsCfVIC-VIC}, where the contingency size is fix. This clearly demonstrates the importance of CC size; hence, accurate identification of CC size would be crucial for frequency stability assessment of FPSs.  
\subsubsection{The Impact of Contingency Location}
\par
\begin{figure}
	\begin{center}
		\includegraphics[width=7.5cm, keepaspectratio]{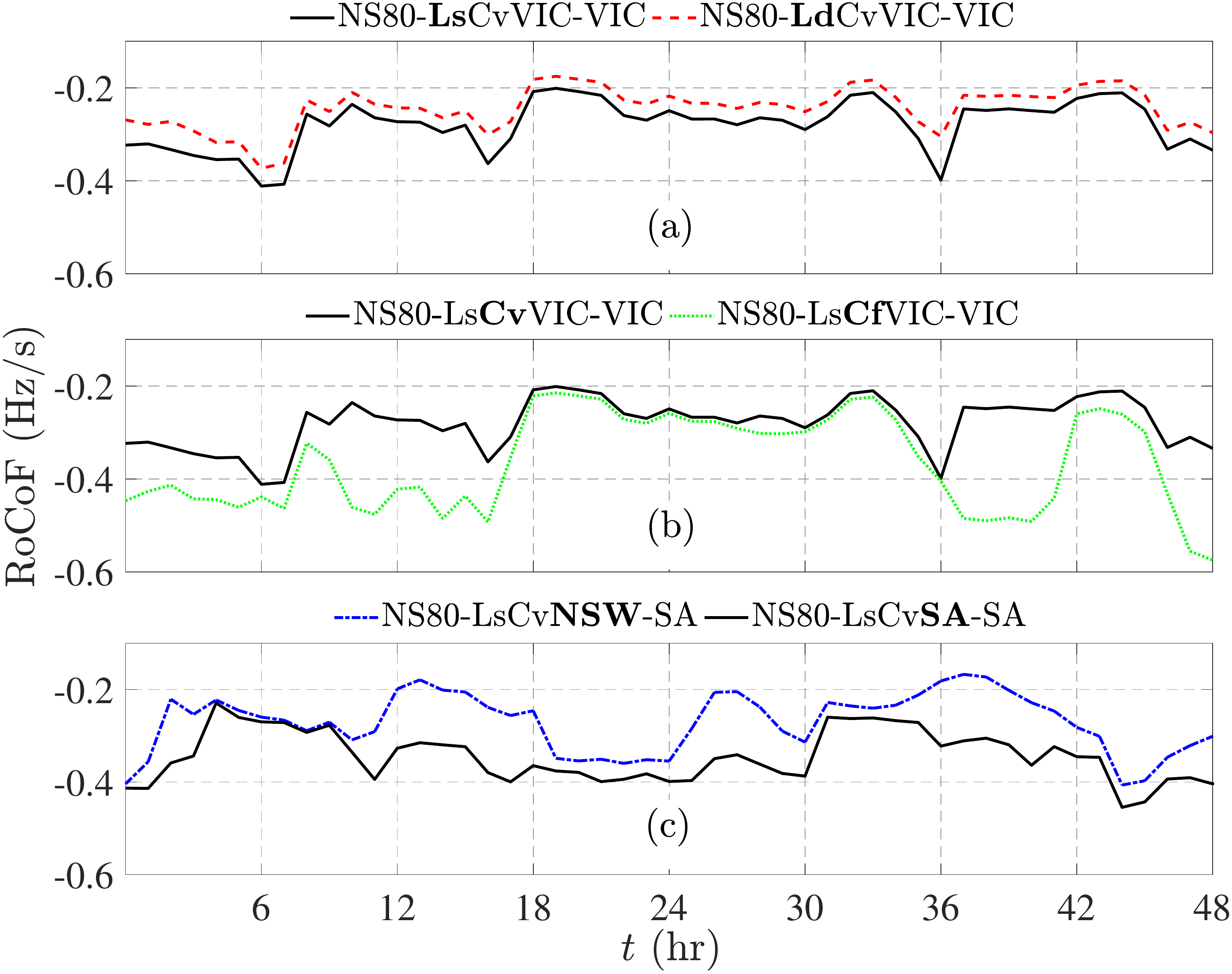}
\vspace{-1.2em}
	\end{center}
	\caption{RoCoF for two typical days considering sensitivity of: (a) load model, (b) CC size, and (c) CC location.}
\vspace{-1.8em}
	\label{LdLs_CfCv_CLocation}
\end{figure}

 The effect of contingency location on RoCoF is shown in Fig.~\ref{LdLs_CfCv_CLocation}c. Observe that in Case~\emph{NS80-LsCvNSW-SA}, where the contingency is occurring in NSW, the RoCoF is not as severe as the Case~\emph{NS80-LsCvSA-SA}, where the incident is in SA. This illustrates the impact of network strength on frequency stability performance. Overall, it was observed that the network strength plays a crucial role on frequency stability behaviour. For instance, the end regions of the NEM that are weakly connected to the system, QLD and SA, were more vulnerable to an incident than the middle regions that are strongly connected to system, NSW and VIC.
\subsubsection{The Impact of Prosumers}
\par
\begin{figure}
	\begin{center}
		\includegraphics[width=7.5cm, keepaspectratio]{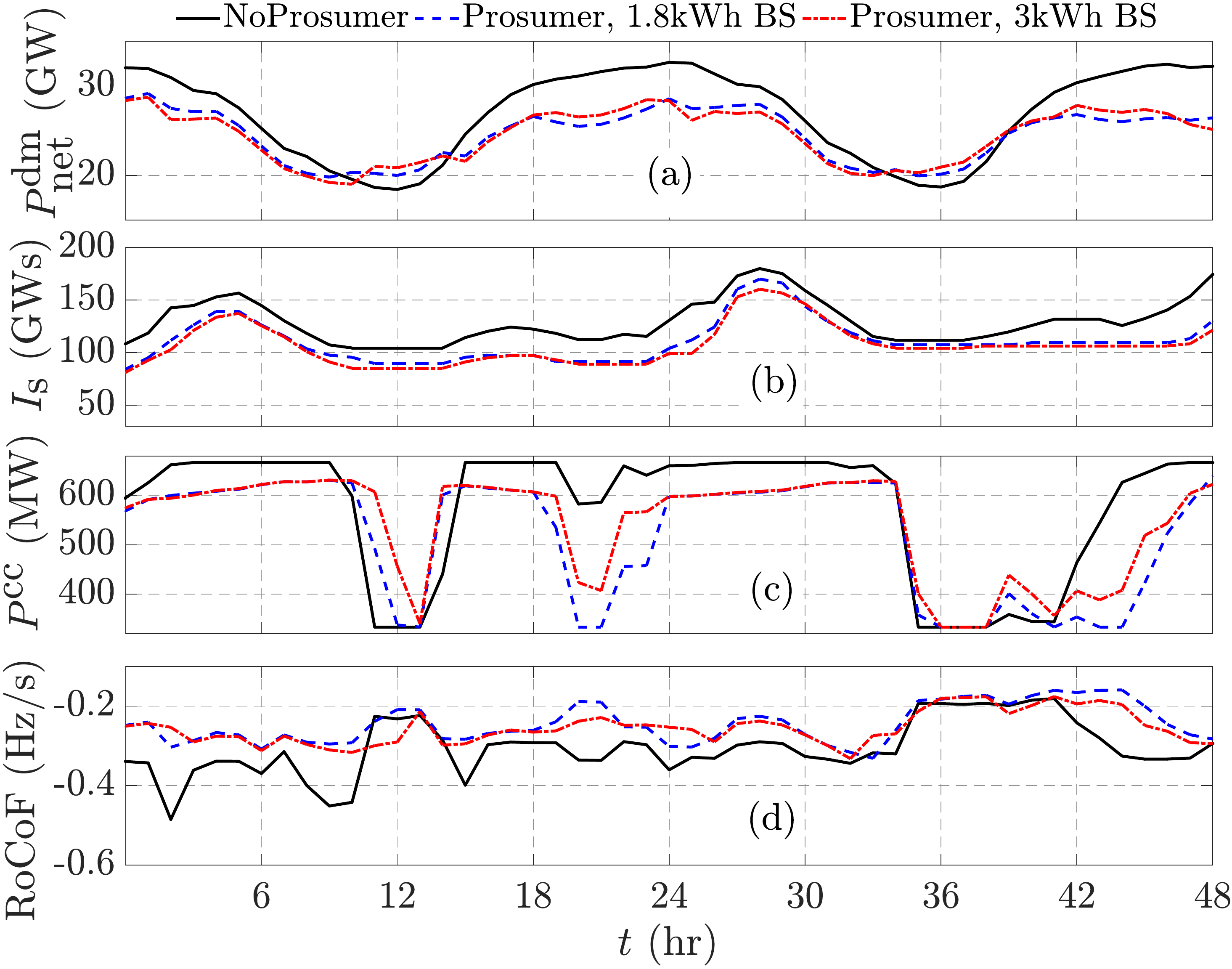}
\vspace{-1.2em}
	\end{center}
	\caption{The impact of prosumers with different BS capacity on: (a) net demand level~($P^{\text{dm}}_{\text{net}}$), (b) level of system inertia ($I_\text{s}$), (c) size of CC~($p^{\text{cc}}$), and (d) RoCoF for two typical days for Case \emph{NS80-LsCvVIC-VIC}.}
	\label{RoCoF_Pro}
\end{figure}

 To study the impact of prosumers on system frequency stability, we use the generic demand model proposed in~\cite{Marzooghi2016ab}, This model aims to capture the aggregated effect of prosumers for market modelling and stability studies. Notice that of the total demand on each state, the percentage of prosumers  with battery storage~(BS) for NSW, VIC, QLD, and SA are \SI{6.3}{\percent}, \SI{8.6}{\percent}, \SI{16}{\percent}, and \SI{22}{\percent}, respectively inspired by AEMO~\cite{AEMO2015b}. Further, to assess the sensitivity of BS capacity, BS capacities of~\SI{1.8}{\kilo\watt\hour}~\cite{AEMO2015b}, and~\SI{3}{\kilo\watt\hour} per ~\SI{1}{\kilo\watt} of rooftop-PV are considered. Observe that with the increased penetration of prosumers, the net demand profile of the system becomes smoother as shown in Fig.~\ref{RoCoF_Pro}a. This is because prosumers utilise their rooftop-PV and exploit the flexibility of their BS system to shift their consumption from peak hours to non-peak hours by maximising their self-consumption~\cite{Marzooghi2016ab}. This, in turn, has an implication on the amount of available inertia as well as the size of CC as shown in Fig.~\ref{RoCoF_Pro}b and c, respectively. Notice that with high penetration of prosumers not only the net demand of the system, but also the system inertia and the size of CC reduces in some hours, e.g. $t=\SI{18}{\hour}~\text{to}~\SI{24}{\hour}$. The combined reduction of system inertia as well as CC size results in a better system RoCoF as illustrated in Fig.~\ref{RoCoF_Pro}d. Overall, it was observed that BS cases had the least number of hours that the RoCoF violated its limits.     
\begin{figure}
	\begin{center}
		\includegraphics[width=7.5cm, keepaspectratio]{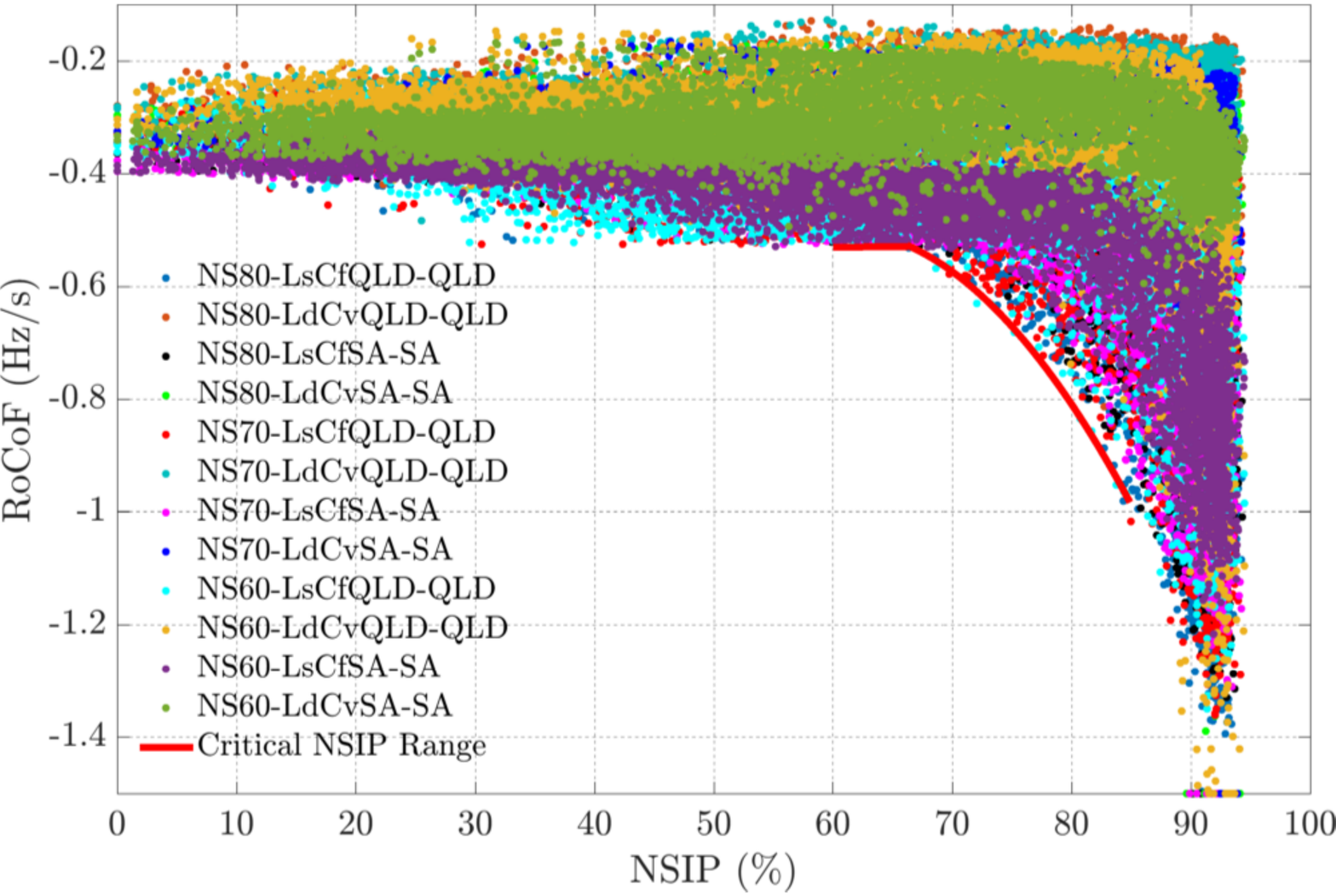}
\vspace{-1.2em}
	\end{center}
	\caption{Minimum RoCoF following a credible contingency based on NSIP.}
\vspace{-1.2em}
	\label{RoCoF_Vs_NSR}
\end{figure}
\vspace{-0.2em}
\subsubsection{Summary}
\par The results suggest that the system frequency response is sensitive to many parameters that can change from one period to the following, which makes it difficult to define a single NSIP that would ensure the stable operation of the system. Therefore, instead of defining a single critical NSIP, a critical NSIP range has to be defined as shown in Fig.~\ref{RoCoF_Vs_NSR}. After analysing the sensitivity of different parameters, we identify the most vulnerable cases, i.e. those with the highest number of hours that the RoCoF violates its limits, and present the results in Fig.~\ref{RoCoF_Vs_NSR}. Observe that for a critical RoCoF, $\frac{df_\text{crt}}{dt}=\SI[]{-0.5}{\hertz\per\second}$, the system critical NSIP range would be \SI{60}{\percent} to \SI{67}{\percent}, which can increase up to \SI{84}{\percent} for a critical RoCoF, $\frac{df_\text{crt}}{dt}=\SI[]{-1}{\hertz\per\second}$. To ensure that the RoCoF is within its permissible bounds in all operation conditions, we will introduce a dynamic inertia constraint in the next Section.           
\vspace{-1.8em}
\section{Improving Frequency Response of the System with High Penetration of NS-RES} \label{Slution}
\par To improve frequency response of the system with high penetration of NS-RES following a contingency, we explore following approaches: 1) adding a dynamic inertia constraint to the market dispatch model, 2) using other resources, such as WFs and SCs.  
\vspace{-1.5em}   
\subsection{Dynamic Inertia Constraint}\label{IC-Constraint}
\par Initially, we implemented a global inertia constraint for the whole system; however, for the NEM, which has a very long transmission network~(over \SI{5000}{\kilo\meter} from QLD to SA), as shown in Fig.~\ref{SLD}, the global inertia constraint was ineffective because inertia location is also very important as demonstrated in Section~\ref{stability_Results}. Therefore, in the time domain simulations, the RoCoF violated its limits due to uneven distribution of inertia in different regions. To assure that each region has sufficient amount of inertia that would prevent the RoCoF from violating its limits, we consider regional RoCoF constraints for each region. To do this, first we identify the size of CC, $p_{r,t}^\text{cc}$, in each region as follows:
\begin{equation}
p_{r,t}^\text{cc}=\mathop{\operatorname{max}}\,(p_{g,t}) \quad g\in \{\mathcal{G^\text{SG}} \cap \mathcal{G}_r \} \text{,}	\label{eq:cntgcy_IC}
\end{equation}
where $\mathcal{G}_r$ is set of all generators in region $r \in \mathcal{R}$. Then, we incorporate the regional inertia constraint in the UC problem. Note that at the inception of an incident, we can assume that the contribution of the load damping on the RoCoF is insignificant; thus, by simplifying equation~(\ref{rocof_formula}), we can consider the following inertia constraint for each region:
\begin{equation} \label{Equation_IC}
I_{\text{s},r} \geq \frac{f_0 p_{r,t}^\text{cc}}{2 |\frac{df_\text{crt}}{dt}|} \text{,} 
\end{equation}
where $I_{\text{s},r}$ represents the total inertia of region $r \in \mathcal{R}$, which is determined considering status~(i.e. on/off), inertia constant $H_g$ and MVA rating of each SG, as given in equation~(\ref{Equation_Inetia_IC}), with $N_r$ being the number of total SG in region $r \in \mathcal{R}$. $f_0$ is system frequency at the inception of the event, $p_{r,t}^\text{cc}$ is the size of CC in region $r \in \mathcal{R}$ identified from equation~(\ref{eq:cntgcy_IC}), , and $\frac{df_\text{crt}}{dt}$ is the critical RoCoF~(i.e. considered as~$\SI[]{-0.5}{\hertz\per\second}$).
\begin{figure}
	\begin{center}
		\includegraphics[width=7.5cm, keepaspectratio]{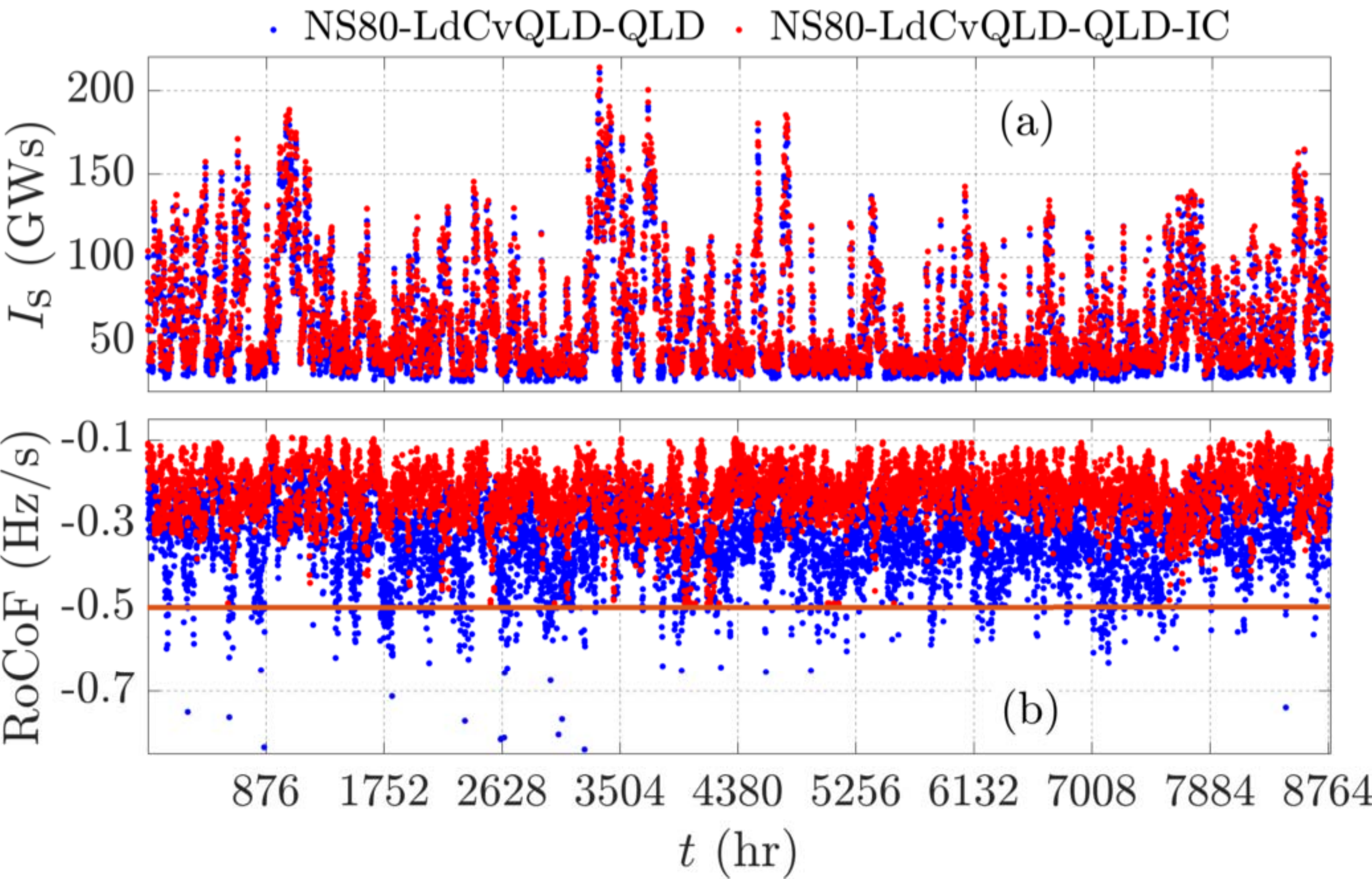}
		\vspace{-1.5em}
	\end{center}
	\caption{Impact of dynamic inertia constraint on: (a) synchronous inertia of the system, and (b) RoCoF for QLD.}
	\vspace{-1.5em}
	\label{KE_RoCoF_Reg_IC}
\end{figure}
\begin{equation} \label{Equation_Inetia_IC}
I_{\text{s},r} = \sum\limits_{i=1}^{N_r} H_{\text{g},i} S_{\text{B},i}. 
\end{equation} 
\par The effectiveness of regional inertia constraints on system synchronous inertia level as well as RoCoF is illustrated in Fig.~\ref{KE_RoCoF_Reg_IC}. Observe that with no inertia constraint in Case~\emph{NS80-LdCvQLD-QLD} the system synchronous inertia can reduce to below $\SI{26}{\giga\watt\second}$. Whereas, with the inertia constraint in Case~\emph{NS80-LdCvQLD-QLD-IC} a minimum synchronous inertia of $I_\text{s}\approx \SI{30}{\giga\watt\second}$ is maintained. As a result, with the inertia constraint, the RoCoF does not violate its limits as shown in Fig.~\ref{KE_RoCoF_Reg_IC}b. Nonetheless, incorporation of inertia constraint in the UC problem results in curtailment of~\SI{380}{\giga\watt\hour} energy from WFs. Note that by curtailing wind energy in a coordinated way, we might be able to add more flexibility to the system, which will be explored in the next subsection. 
\begin{figure}
	\begin{center}
		\includegraphics[width=7.5cm, keepaspectratio]{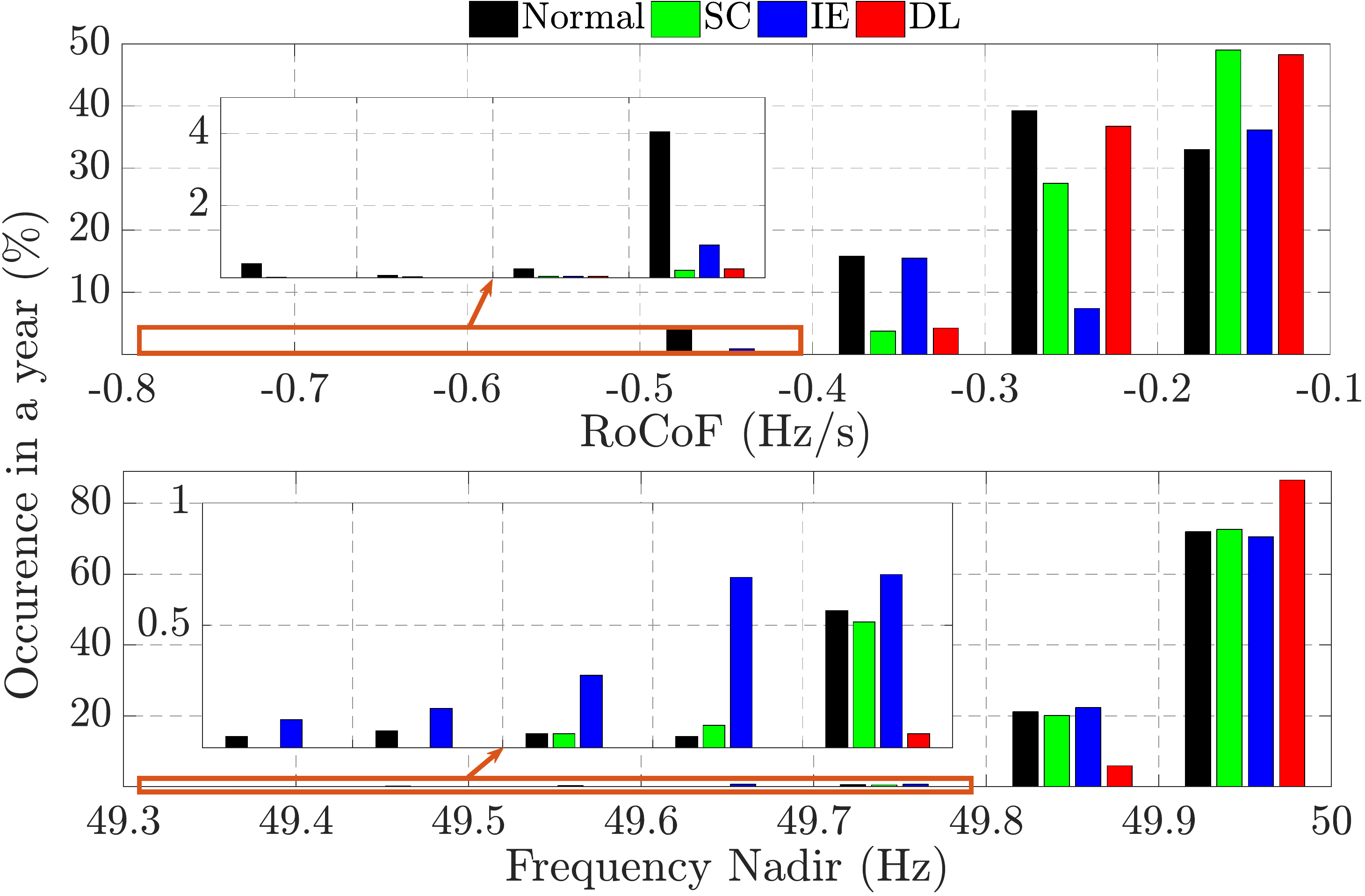}
\vspace{-1.5em}
	\end{center}
	\caption{Distribution of: (a) RoCoF, and (b) frequency nadir for Case~\emph{NS80-LdCvQLDQLD}; considering~\textbf{Normal} operation, \textbf{SC}, \textbf{IE}, and \textbf{DL}.}
\vspace{-1.2em} 
	\label{Df_dfdt_SC_DFIG}
\end{figure}
\begin{figure}
	\begin{center}
		\includegraphics[width=7.5cm, keepaspectratio]{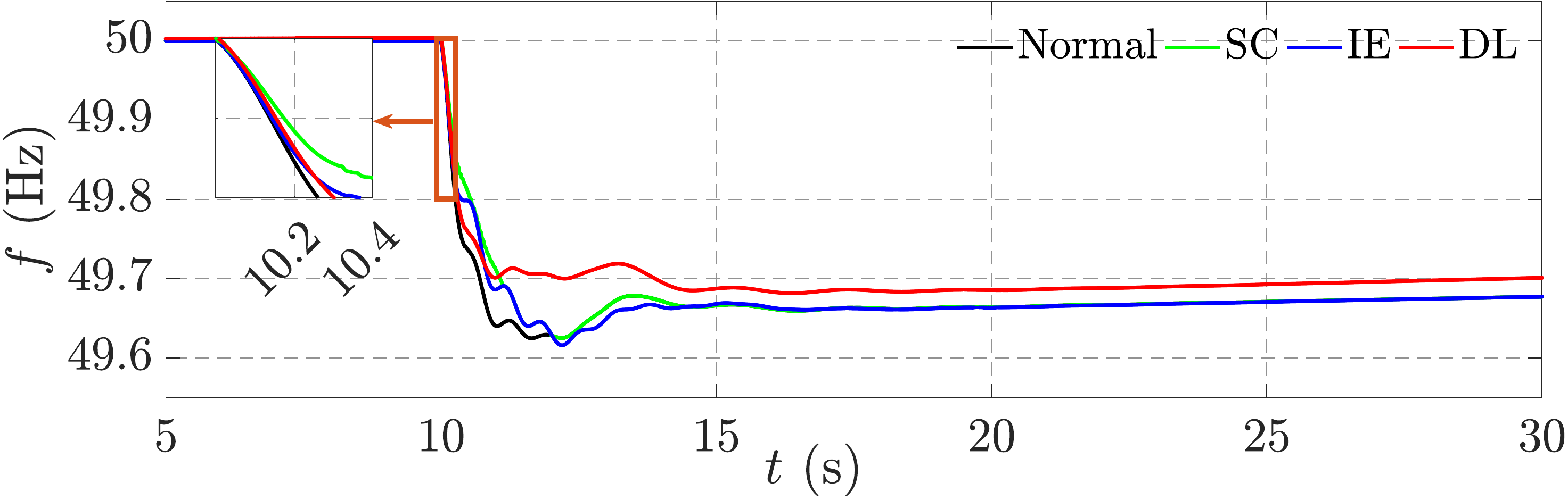}
\vspace{-1.5em}
	\end{center}
	\caption{System frequency behaviour for a typical hour considering different options, i.e.~\textbf{Normal} operation, \textbf{SC}, \textbf{IE}, and \textbf{DL}.}
\vspace{-1.8em}
	\label{Time_Domain_Fre_Feg}
\end{figure}
\vspace{-0.8em}
\subsection{Utilisation of Other Sources for Frequency Control} \label{Improving_RoCoF_WF_SC}
\par To improve system frequency response, we use three different techniques and perform time-series analysis for a whole year. This will allow us to compare and quantify the contribution of each option. To do this, we use Case~\emph{NS80-LdCvQLD-QLD} as the base case and explore the following options: 
\begin{itemize}
\item \textbf{Normal}: This is the reference option and no additional source is used for improving system frequency response.
\item \textbf{SC}: In every region, we consider one synchronous condenser~(SC), with ($S=\SI{400}{\mega\volt\ampere}$) and ($H=\SI{6}{\second}$). This would add a total synchronous inertia of ($I_\text{s}=\SI{9.6}{\giga\watt\second}$) to the system around the clock.
\item \textbf{IE}: Since the contingency is located in QLD, we consider one of the QLD's WFs that operates at near its rated capacity and provides $P_{\text{WF},t}=\SI{600}{\mega\watt}$ all over the year. We apply the inertia emulation~(IE) technique~\cite{Morren2006} to exploit the rotational kinetic energy of this WF for frequency control. 
\item \textbf{DL}: We use the same WF introduced earlier and de-load~(DL) it by \SI{5}{\percent}, which results in $P_{\text{WF},t}=\SI{570}{\mega\watt}$. The de-loading is achieved by pitching the blade angles and increasing the rotor speeds of the wind turbines~\cite{Ahmadyar2017}. Note that the total curtailed energy of the WF over the year would be~\SI{263}{\giga\watt\hour} compared to \SI{380}{\giga\watt\hour} resulted from implementation of inertia constraint.  
\end{itemize} 
\par Fig.~\ref{Df_dfdt_SC_DFIG} summarises the minimum RoCoF and the frequency nadir of the system for all options. Observe that in the \emph{Normal} option the RoCoF violates its critical value~(i.e. $\SI[]{-0.5}{\hertz\per\second}$) in many hours; whereas, in the other options, i.e. \emph{SC}, \emph{IE} and \emph{DL}, the RoCoF does not violate its critical value and falls between $\SI[]{-0.5}{\hertz\per\second}$ and $\SI[]{-0.1}{\hertz\per\second}$. Thus, from the RoCoF perspective, the most effective option is \emph{SC} followed by \emph{DL} and \emph{IE}, as shown in Fig.~\ref{Time_Domain_Fre_Feg}. This is because in Option~\emph{SC} the amount of synchronous inertia is considerably high compared to other options. Further, in contrast to \emph{DL} and \emph{IE} options that require control action to provide inertial response, \emph{SC} provides natural response, which is more effective. Note that the available kinetic energy of the WF in Option~\emph{DL} is more than Option~\emph{IE} because the de-loading is achieved by increasing the rotor speed of the wind turbines as well as pitching of the wind turbines' blades. 
\par The frequency nadir, however, has been affected differently as shown in Fig.~\ref{Df_dfdt_SC_DFIG}b. For instance, in Option~\emph{SC}, since the SC provides inertial response but not governor response, it reduces the minimum RoCoF and delays the frequency nadir, but has insignificant impact on the value of the frequency nadir. In Option~\emph{IE}, the RoCoF improves; however, the frequency nadir deteriorates compared to base case in some hours as shown in~Fig.~\ref{Df_dfdt_SC_DFIG}b. This is because extraction of kinetic energy from a wind turbine reduces its rotor speed, and in turn, its coefficient of performance. Recovery of the coefficient of performance following inertia emulation requires kinetic energy that should be provided by the grid. In some situations, this would result in a second frequency nadir as shown in Fig.~\ref{Time_Domain_Fre_Feg}. As a result, the number of hours that the frequency nadir falls between \SI{49.3}{\hertz} and \SI{49.8}{\hertz} slightly increases compared to base case. Nonetheless, Option~\emph{DL} not only improves RoCoF and frequency nadir, but it also has a positive impact on settling frequency; because, in addition to kinetic energy, it releases additional power to the grid using a governor-like control.    
\vspace{-0.8em}
\section{Conclusions} \label{concl}
\par This paper introduces a framework for frequency stability assessment of future power systems. By utilising this framework, we performed a comprehensive scenario based sensitivity analysis on the Australian future power system, and identified a critical range of non-synchronous instantaneous penetration that the system can accommodate without violating its frequency stability limits. Further, it was shown that there are many parameters that affect the frequency stability of the system that change from one period to the following. Therefore, to consider those changes, and to maintain system frequency response within its permissible bounds at all time, we proposed a dynamic inertia constraint  and incorporated it into the market dispatch model. It was shown that by implementation of such constraint the frequency control criteria can be satisfied. Nonetheless, enforcement of this constraint resulted in curtailment of wind energy. We showed that by curtailing wind energy in a coordinated way we can add more flexibility to the system. This was done by de-loading the wind farms and engaging them to frequency control using a governor-like response, which not only improved the RoCoF and the frequency nadir, but also improved the settling frequency following a contingency. Moreover, by performing extensive time-series simulation, we quantified the contribution of synchronous condenser and wind farm synthetic inertia on system frequency response.  

\ifCLASSOPTIONcaptionsoff
  \newpage
\fi



%
\vspace{-0.6em}
\bibliographystyle{IEEEtran}
\bibliography{IEEEJounralone}

%







\end{document}